\documentclass[lettersize,journal]{IEEEtran}
\usepackage{amsmath,amsfonts}
\usepackage{algorithmic}
\usepackage{array}
\usepackage[caption=false,font=normalsize,labelfont=sf,textfont=sf]{subfig}
\usepackage{textcomp}
\usepackage{stfloats}
\usepackage{url}
\usepackage{verbatim}
\usepackage{graphicx}
\def\BibTeX{{\rm B\kern-.05em{\sc i\kern-.025em b}\kern-.08em
    T\kern-.1667em\lower.7ex\hbox{E}\kern-.125emX}}
\usepackage{balance}

\usepackage[T1]{fontenc}
\usepackage{bm} 
\usepackage{caption} 
\usepackage{cite} 
\usepackage{algorithmic} 
\usepackage{algorithm} 
\usepackage{multirow} 
\usepackage{graphicx} 
\usepackage{url} 
\begin{document}

\bstctlcite{setting}
\title{Segmented Exponent Alignment and Dynamic Wordline Activation for Floating-Point Analog CIM Macros
}
\author{Weiping Yang, Shilin Zhou, Hui Xu, Jiawei Xue, Changlin Chen\(^*\)\\
\textit{National University of Defense Technology, College of Electronic Science and Technology, Changsha, China}\\
{email:\(^*\)changlinchen@nudt.edu.cn,  yangwp2021@nudt.edu.cn}

\vspace{-0.3cm}
\thanks{This work was supported in part by the Hunan Provincial Innovation Foundation for Postgraduate under Grant XJZH2024008. 

}
}

\markboth{
\textbf{A\MakeLowercase{ccepted version for }IEEE ICECS 2025, M\MakeLowercase{arrakech}, M\MakeLowercase{orocco}. }
\MakeLowercase{\textcopyright~2025 Copyright held by the authors. Publication rights licensed to} IEEE. 
DOI:\MakeLowercase{https://doi.org/10.1109}/ICECS/XXX.
}{}

\maketitle
\begin{abstract}
With the rise of compute-in-memory (CIM) accelerators, floating-point multiply-and-accumulate (FP-MAC) operations have gained extensive attention for their higher accuracy over integer MACs in neural networks.
However, the hardware overhead caused by exponent comparison and mantissa alignment, along with the delay introduced by bit-serial input methods, remains a hinder to implement FP-MAC efficiently.
In view of this, we propose Segmented Exponent Alignment (SEA) and Dynamic Wordline Activation (DWA) strategies.
SEA exploits the observation that input exponents are often clustered around zero or within a narrow range. By segmenting the exponent space and aligning mantissas accordingly, SEA eliminates the need for maximum exponent detection and reduces input mantissa shifting, and thus reduces the processing latency. DWA further reduces latency and maintains accuracy by activating wordlines based on the exponent segments defined by SEA.
Simulation results demonstrate that, when compared with conventional comparison tree based maximum exponent alignment method, our approach saves 63.8\% power consumption, and achieves a 40.87\% delay reduction on the VGG16-CIFAR10 benchmark.
\end{abstract}

\vspace{-0.8cm}
\section{Introduction}
\vspace{-0.3cm}
Conventional compute-in-memory (CIM) macros often employ integer-based  multiply-and-accumulate (INT-MAC) operations for the sake of simplicity \cite{mengjssc23,aspadc24,rramTCASI21,heirsDAC24,pipecim}. However, large language models and complex neural network tasks require floating-point multiply-and-accumulate (FP-MAC) operations, since FP-MACs offer higher accuracy than their integer-based counterparts \cite{neugpu24}. However, floating-point CIM (FP-CIM) macros remain underexplored due to their inherently complex computation processes \cite{16n96k24,floatCIMiscas22,mengisscc24,tufengbin22}.

\begin{figure}[H]
\centering
\includegraphics[width=9cm]{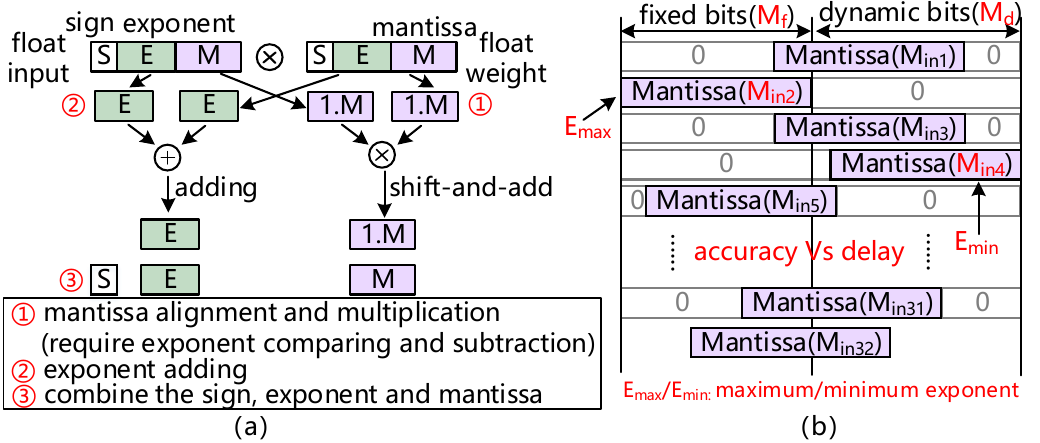}
\caption{(a) FP-MAC procedure and (b) Mantissa alignment}
\label{sem2}
\end{figure}

Different from integer-based CIMs which require only MAC operations, FP-CIMs requires extra operations like exponent comparison and mantissa alignment.
Fig.~\ref{sem2}(a) illustrates the three steps to implement floating-point (FP) computation. The mantissas need to be shifted and aligned according to the maximum exponent. The FP-CIM input width must be expanded from the standard fixed mantissa width ($\mathrm{M_f}$) to a dynamic width ($\mathrm{M_f + M_d}$) determined by the exponent difference, as shown in Fig.~\ref{sem2}(b).

Previous FP-CIM solutions attempt to simplify the computation process by introducing alternative data formats,
such as the block floating-point (BF) format \cite{bfpcim24}, microscaling format \cite{khwa25}, POSIT format \cite{positJSSC25}, and logarithmic-based format \cite{log24jssc}. However, these formats require customized training and have limited application scenrio, unlike the widely adopted IEEE-754 FP16 standard \cite{ieee754}.
\vspace{-0.2cm}
Another approach involves using heterogeneous digital macros to handle exponent mantissas operations. S. Yan et al. \cite{syan24} notice that most exponent values in floating-point data are concentrated in a narrow range. They propose a hybrid architecture that combines intensive CIM macros with sparse digital cores: dense inputs and dense weights are handled by CIM macros, while all other computations are offloaded to digital core. The heterogeneous architecture demands careful categorization and distribution of inputs and weights across different macros, requiring complex scheduling circuits. The sparse encoding techniques also introduce extra overhead.

In analog FP-CIM design, H. Hsu et al. \cite{resADC25} proposed a rescheduled input (RI) scheme that compresses 2-bit inputs into 1-bit with a flag, cutting input cycles and reducing latency by about 50\%. However, this method incurs high hardware overhead and does not optimize input exponent operations.

In summary, despite prior progress, opportunities still remain to optimize input exponent and mantissa handling in FP-CIM architectures. In view of this, we propose an analog CIM macro incorporating Segmented Exponent Alignment (SEA) and Dynamic Wordline Activation (DWA) strategies to reduce exponent-processing overhead and input latency.

Inspired by S. Yan et al. \cite{syan24}, we analyzed exponent distributions and found that most FP16 exponents are not only concentrated in a narrow range but are also frequently clustered around zero. SEA exploits this by dividing the exponent space into three regions based on the 3 MSBs, avoiding global maximum exponent detection and greatly reducing mantissa shifting lengths and accuracy loss. DWA supports SEA by activating wordlines for each exponent group in separate cycles, avoiding complex dual type macro designs and sparse encodings in \cite{syan24}. SEA and DWA preserve more mantissa bits and provides shorter input latency. Evaluations on the VGG16-CIFAR10 benchmark demonstrate a 40.87\% reduction in bit-serial delay with only a 2.04\% accuracy loss.

\section{Proposed Strategy}
\begin{figure}[H]
\centering
\includegraphics[width=9cm]{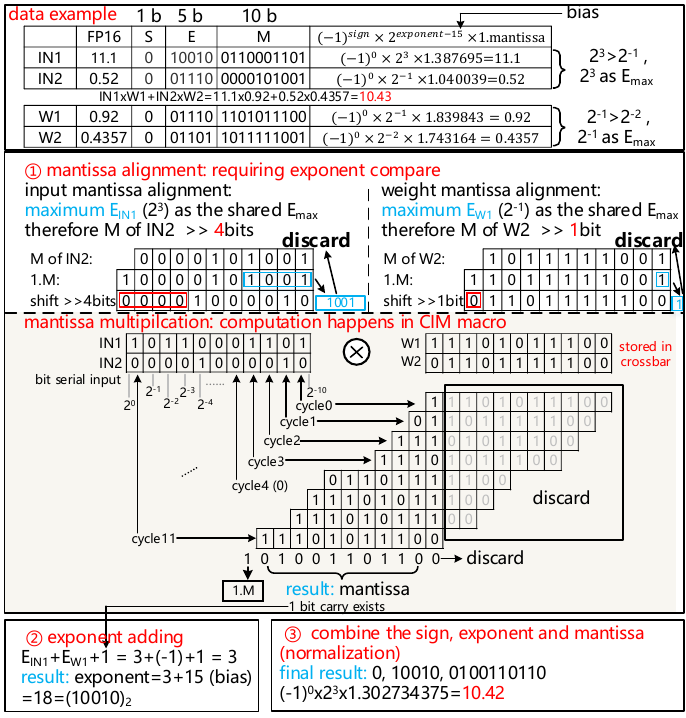}
\caption{Illustration of FP computation process and Fixed Width Input (FWI) method in FP-CIM}
\label{eg}
\end{figure}

Conventionally, the pre-aligned mantissas with a width of $\mathrm{M_f + M_d}$ in Fig.~\ref{sem2}(b) are fed bit-serially into the CIM macro. We refer to this method as the Dynamic Width Input (DWI) method, which results in high input latency. 
To mitigate this, the Fixed Width Input (FWI) method retains only a fixed number of bits ($\mathrm{M_f}$), significantly shortening bit-serial cycles. However, FWI may degrade accuracy due to the loss of lower-bit dynamic information.
Fig.~\ref{eg} presents an example of FWI-based FP16 computation process, illustrating $IN1 \times W1 + IN2 \times W2$. In FP16, each value is expressed as $(-1)^{sign} \times 2^{exponent - 15} \times 1.mantissa$, where 15 is the exponent bias.

\vspace{-0.4cm}
\vspace{-0.4cm}

Fig.~\ref{sem2}(b) and Fig.~\ref{eg} also show that large exponent differences within an input group can cause mantissas with smaller exponents, such as \(\mathrm{M_{in4}}\) in Fig.~\ref{sem2}(b), to be dropped during shifting. This leads to accuracy loss, especially when inputs exponent distributions are highly unbalanced.

Moreover, the weights in FP-CIMs are pre-aligned offline and stored in mantissa form within the CIM crossbar, while inputs generated dynamically require on-chip alignment, which is typically performed by a comparison tree \cite{tufengbin22, apccas22}. However, utilizing comparison-tree to detect the maximum exponent within an input group is energy-intensive\cite{apccas22}.

Based on the above analysis, we propose the SEA strategy to simplify input exponent comparison. The SEA strategy reduces mantissa shift distances by avoiding alignment to the global maximum exponent. Instead, SEA segments the input exponent space and aligning input mantissas accordingly. Different from S. Yan et al. \cite{syan24} who adopt dual macro to deal with the intensive and sparse input which requires sparse encoding, we introduce a DWA strategy to activate different input groups in different cycles with slight overhead. Combined with FWI, our proposed SEA and DWA strategies mitigates bit loss due to excessive right shifts.

\vspace{-0.3cm}
\subsection{Segmented Exponent Alignment}
\vspace{-0.4cm}
\begin{figure}[H]
\centering
\includegraphics[height=6cm]{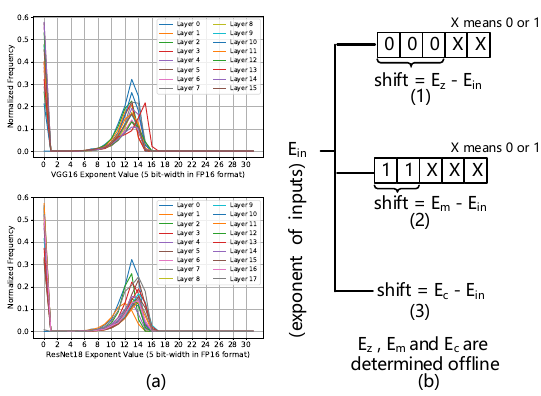}
\vspace{-0.3cm}
\caption{(a) Distribution of input exponents in VGG16-CIFAR10 \& ResNet18-CIFAR10 (b) Proposed 3 MSB-based categories }
\label{layerid}
\end{figure}

The distribution statistics of input exponents for each layer in VGG16-CIFAR10\cite{vgg16} and ResNet18-CIFAR10\cite{resnet18cvpr} are illustrated in Fig \ref{layerid}(a). 
We can observe that input exponents are concentrated within two separate ranges. Based on this distribution, exponent values can be segmented into several categories. Specifically, we classify the input exponents into three groups: near-zero, center, and near-maximum. The corresponding mantissa shifting operations can also be simplified.

\begin{figure}[H]
\centering
\includegraphics[width=8cm]{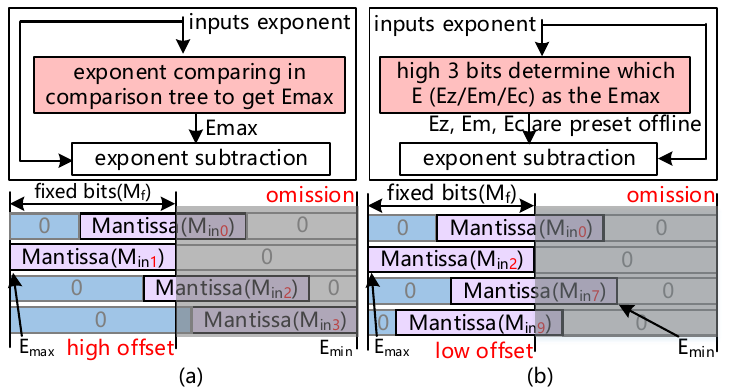}
\caption{(a) Conventional strategy with FWI (b) Proposed SEA and DWA strategies with FWI}
\label{vs}
\end{figure}

Specifically, given a 5-bit input exponent \( \mathrm{E_{in}} \), the highest 3 bits (3 MSBs) are examined to determine the exponent category. If the 3 MSBs are allo zero, the exponent is considered in the near-zero region, and the alignment shift is computed as equation (1) in Fig.~\ref{layerid}(b), where \( \mathrm{E_z} \) is a predefined shared exponent for the near-zero region. If 3 MSBs are 3'b11X, the exponent is classified as belonging to the near-maximum region, and the shift is given by equation (2) in Fig.~\ref{layerid}(b), where \( \mathrm{E_m} \) represents the shared exponent for large values. In all other cases, the exponent is considered part of the center region, and the alignment is calculated as equation (3) in Fig.~\ref{layerid}(b), where \( \mathrm{E_c}\) is the center shared exponent. This categorization helps minimize mantissa alignment shifts and adapt to the FWI method, as shown in Fig.~\ref{vs}.

In Fig.~\ref{vs}, the conventional strategy suffers from high offsets when exponent differences are large, causing severe mantissa truncation. In contrast, our method effectively alleviates this problem by minimizing unnecessary shifts. This benefit comes from the proposed DWA strategy, which activates wordlines based on SEA-derived groups rather than in sequential order. The \(\mathrm{Min_{0/2/7/9}}\) are activated in Fig.~\ref{vs}(b) instead of \(\mathrm{Min_{0/1/2/3}}\) in Fig.~\ref{vs}(a).
\vspace{-0.3cm}
\subsection{Dynamic Wordline Activation}
\begin{figure}[H]
\centering
\includegraphics[width=8cm]{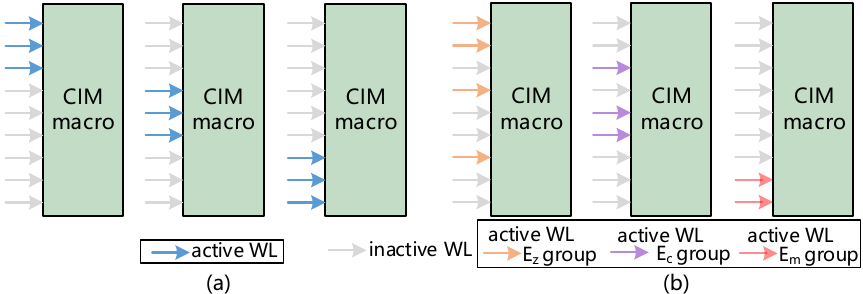}
\caption{(a) In-order wordline (WL) operation (b) Proposed Dynamic Wordline Activation Strategy (DWA)}
\label{wls}
\end{figure}

The proposed Dynamic Wordline Activation Strategy is shown in Fig.~\ref{wls}. In the conventional in-order wordline (WL) operation (Fig.~\ref{wls}(a)), all input WLs are activated sequentially regardless of their exponent characteristics, which leads to excessive mantissa alignment shifts when exponent values vary significantly, as shown in Fig.~\ref{vs} (a). In contrast, the proposed DWA activates WLs based on the exponent category of the input data. Specifically, WLs corresponding to the \(\mathrm{E_z}\) group, \(\mathrm{E_m}\) group and \(\mathrm{E_c}\) group are activated separately, as shown in Fig.~\ref{wls} (b).

The only overhead introduced by DWA is a 2-bit flag per input, indicating its SEA group (Ez, Ec, or Em), which is minimal compared to the dual-macro scheduling and sparse encoding overhead reported in S. Yan et al. \cite{syan24}.
\vspace{-0.3cm}

\vspace{-0.2cm}
\subsection{Integrating SEA and DWA into the FP-CIM}

\vspace{-0.2cm}
\begin{figure}[H]
\centering
\includegraphics[width=8.5cm]{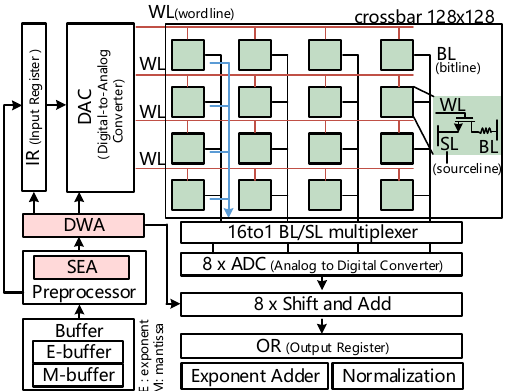}
\vspace{-0.1cm}
\caption{Proposed analog FP-CIM macro architecture}
\label{arch}
\end{figure}
\vspace{-0.2cm}
Our proposed strategies and circuit are implemented in an analog FP-CIM, whose overall architecture is shown in Fig.~\ref{arch}. Before computation, the weight mantissas are pre-aligned and stored in the crossbar during an offline phase.

During computation, input exponents and mantissas are fetched from the E-buffer and M-buffer by the Preprocessor. The exponents are then processed by the SEA module to determine the right-shift amounts and grouping information for the input mantissas. Then  input mantissas are shifted and loaded into the Input Registers (IR), while the grouping information is sent to the DWA module. DWA module controls the IR to sequentially send mantissas to the DAC based on the grouping, thereby activating the corresponding wordlines (WLs) in the crossbar. 

The resulting bitline currents are digitized by the ADC and accumulated over multiple cycles using shift-and-add modules. The partial sums are stored in the Output Register (OR). After mantissa computation, the exponents of inputs and weights are combined using an Exponent Adder. Finally, the Normalizer module merges the exponent and mantissa to produce the final floating-point output. 

Note that the segmentation and dynamic activation approach is flexible and can be tailored to the characteristics of different neural networks.

\vspace{-0.3
cm}
\section{Simulation and Comparison Results}
\vspace{-0.1cm}
To validate the proposed SEA and DWA strategies, we design our FP-CIM in a 28nm process.
The power and area metrics  of the circuits operating at 100MHz are shown in table \ref{ppa_break}. These results are obtained through a combination of Cadence Virtuoso for analog simulation, Synopsys Design Compiler for digital synthesis, and Calibre for physical verification.
\begin{table}[H]
\centering
\caption{Power\&Area of our FP-CIM@28nm,100MHz,1.0V}
\begin{tabular}{|c|c|c|}
\hline
\textbf{Component} & \textbf{Power (\(\mathrm{mW}\))} & \textbf{Area (\(\mathrm{mm^2}\))} \\ \hline
SEA & 0.045132& 0.000312 \\ \hline
Preprocessor w/o SEA & 2.265700& 0.034605\\ \hline
DWA &  0.151200 & 0.001311 \\ \hline

Input Register (IR) & 2.138300& 0.026760 \\ \hline
DAC & 0.001677&  0.007788\\ \hline
16to1 BL/SL Multiplexer & 0.000018& 0.001078 \\ \hline
(SAR) ADC (6 bit) & 0.082214& 0.000494\\ \hline

Shift-and-Add & 0.001840  & 0.000117\\ \hline
Output Register (OR) & 1.081500& 0.012767\\ \hline
Exponent Adder & 0.009175 &   0.000079 \\ \hline
Normalization &  0.304200 & 0.001806\\ \hline
E-buffer & 0.113600  & 0.000916\\ \hline
M-buffer & 0.207500 &  0.001657\\ \hline
\end{tabular}
\label{ppa_break}
\end{table}
\vspace{-0.2cm}

To evaluate the efficiency of the proposed SEA and DWA strategies, we compare them against state-of-the-art designs from F. Tu et al. \cite{tufengbin22}, S. Yan et al. \cite{syan24}, and H. Hsu et al. \cite{resADC25}, as shown in Table~\ref{compre}.

\begin{figure}[H]
\centering
\includegraphics[height=4cm]{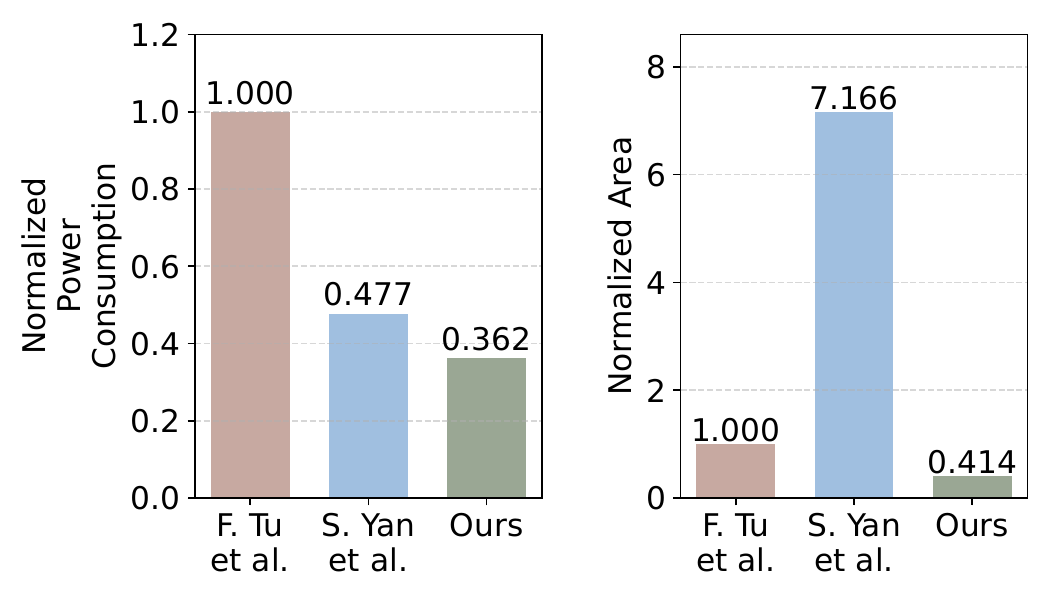}
\vspace{-0.2cm}
\caption{Power \& area profits of our SEA+DWA strategies in 28nm at 100MHz}
\label{fig:ppa0}
\end{figure}

For a fair evaluation of exponent handling and dataflow strategies, we normalize the area and power overhead specific to these mechanisums with respect to the baseline of F. Tu et al. \cite{tufengbin22}, which employs conventional MEA and Booth encoding. As shown in Fig.~\ref{fig:ppa0}, our SEA and DWA techniques reduce power and area by 63.8\% and 58.6\%,  respectively. In contrast, S. Yan et al.'s \cite{syan24} dual-macro design achieves a smaller power (47.7\%) but incurs a 7.17x area overhead due to complex scheduling and encoding logic. H. Hsu et al. \cite{resADC25} provide optimization on latency reduction but does not optimize exponent handling and is excluded from this comparison.

\vspace{-0.2cm}
\begin{table}[H]
\centering
\caption{Comparison Results}
\begin{tabular}{|c|c|c|c|c|}
\hline
\textbf{Works}  & \begin{tabular}[c]{@{}c@{}}\textbf{This} \\ \textbf{work}\end{tabular}
& \begin{tabular}[c]{@{}c@{}}\textbf{S. Yan et al.}\\ \textbf{ (JSSC'24)}\\ \cite{syan24}\end{tabular}
& \begin{tabular}[c]{@{}c@{}}\textbf{H. Hsu et al.}\\ \textbf{(JSSC'25)}\\ \cite{resADC25}\end{tabular}
& \begin{tabular}[c]{@{}c@{}}\textbf{F. Tu et al.}\\ {}\textbf{(ISSCC'22)}\\ \cite{tufengbin22}{}\end{tabular} \\ \hline

\textbf{Process} & 28nm & 28nm & 22nm & 28nm\\ \hline
\begin{tabular}[c]{@{}c@{}}\textbf{CIM} \\ \textbf{Type}\end{tabular}
&  \begin{tabular}[c]{@{}c@{}}Analog \\ (RRAM)\end{tabular}
& \begin{tabular}[c]{@{}c@{}}Digital \\ (SRAM)\end{tabular}
& \begin{tabular}[c]{@{}c@{}}Analog \\ (RRAM)\end{tabular}
&  \begin{tabular}[c]{@{}c@{}}Digital \\ (SRAM)\end{tabular}\\ \hline
\textbf{Data Type} & FP16 & FP16 &  FP/BF16 & BF16/FP32\\ \hline
\begin{tabular}[c]{@{}c@{}}\textbf{Macro} \\ \textbf{Size}\end{tabular}
& \begin{tabular}[c]{@{}c@{}}128x128 \\ (16Kb)\end{tabular}
& \begin{tabular}[c]{@{}c@{}}128x128 \\ (16Kb)\end{tabular}
& \begin{tabular}[c]{@{}c@{}}1024x1024 \\(1Mb)\end{tabular}
&  \begin{tabular}[c]{@{}c@{}}8x48 \\(0.375Kb)\end{tabular}\\ \hline

\begin{tabular}[c]{@{}c@{}} \textbf{Expo. \&} \\ \textbf{Input} \\ \textbf{Strategy}\end{tabular}
& \begin{tabular}[c]{@{}c@{}}SEA+ \\ DWA+\\FWI\end{tabular}
& \begin{tabular}[c]{@{}c@{}} CIM\&digital \\dual macro \\+DWI \end{tabular}
& \begin{tabular}[c]{@{}c@{}} 2 to 1 bits\\ encoding\\ with a flag\\ \end{tabular}
&  \begin{tabular}[c]{@{}c@{}}Comparison\\ tree+Booth \\encoding+\\DWI \end{tabular} \\ \hline

\begin{tabular}[c]{@{}c@{}}\textbf{Power} \\ \textbf{(mW)}\end{tabular}
& \begin{tabular}[c]{@{}c@{}}0.196332 \end{tabular}
& \begin{tabular}[c]{@{}c@{}}0.258982\end{tabular}
& \begin{tabular}[c]{@{}c@{}} NaN\end{tabular}
&  \begin{tabular}[c]{@{}c@{}} 0.542600\end{tabular}\\ \hline

\begin{tabular}[c]{@{}c@{}}\textbf{Area} \\ \textbf{(}\(\mathrm{\mathbf{\mu m^2}}\)\textbf{)}\end{tabular}
& \begin{tabular}[c]{@{}c@{}}0.001623 \end{tabular}
& \begin{tabular}[c]{@{}c@{}}0.028083\end{tabular}
& \begin{tabular}[c]{@{}c@{}} NaN\end{tabular}
&  \begin{tabular}[c]{@{}c@{}} 0.003919\end{tabular}\\ \hline

\end{tabular}
\label{compre}
\end{table}

\vspace{-0.4cm}
\begin{figure}[H]
\centering
\includegraphics[width=9cm]{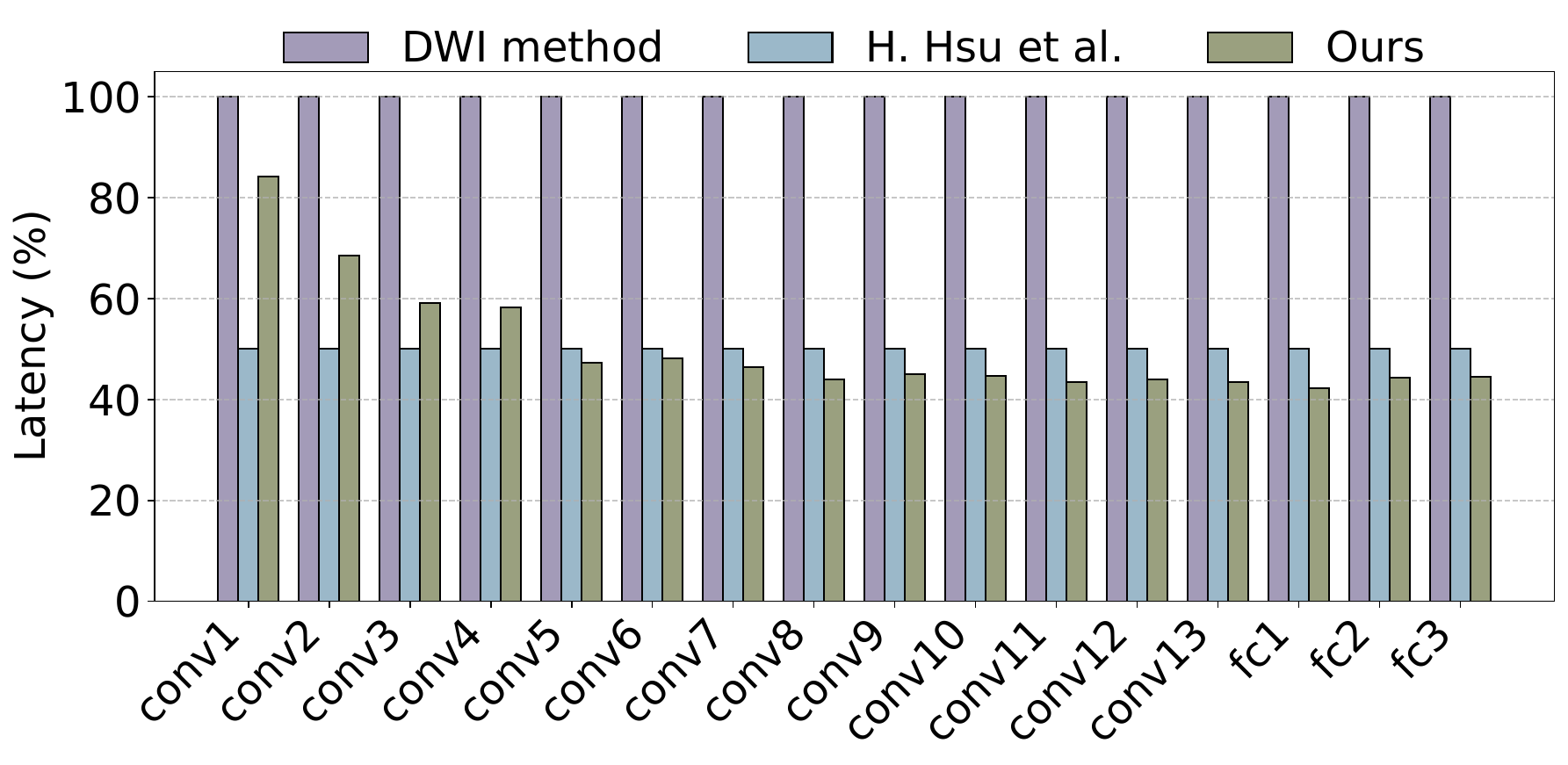}
\caption{Layerwise reduced latency in VGG16-CIFAR10}
\label{vgg16layer}
\end{figure}

\vspace{-0.2cm}
Owing to the adoption of an FWI-based design rather than a DWI-based one, our approach offers a notable reduction in input latency. We evaluate bit-serial input latency reduction on the VGG16-CIFAR10 task by counting the bit-serial input cycles. As shown in  Fig.~\ref{vgg16layer}, our layerwise analysis demonstrates an average 40.87\% latency reduction with only 0.15mW overhead, significantly outperforming the DWI scheme used by S. Yan et al. \cite{syan24} and F. Tu et al \cite{tufengbin22}, across both convolutional and fully connected layers. While H. Hsu et al.'s approach \cite{resADC25} achieves roughly 50\% of DWI's latency, it has over 5.5x higher power overhead than ours. Experiments on ResNet18-CIFAR10 show a similar effect.

\vspace{-0.3cm}

Regarding accuracy, we compare our SEA+DWA \& FWI scheme with the SEA+DWA \& DWI scheme. Simulation results show that our design incurs only a 2.04\% loss on VGG16-CIFAR10 task.

\section{Conclusion}
In this paper, we propose a Segmented Exponent Alignment (SEA) strategy that eliminates the need to detect the maximum exponent, to exploit the benefits that exponent values are typically concentrated within a small range or near zero. To support the SEA strategy, we integrate the Dynamic Wordline Activation (DWA) strategy into an analog CIM, effectively reducing bit-serial cycles. Simulations on the VGG16-CIFAR10 task demonstrate an average latency reduction of 40.87\%.


\end{document}